\documentclass[12pt,preprint]{aastex}
\usepackage{amsmath}
\usepackage{amssymb}
\usepackage{graphicx}

\newcommand\bmath[1] {\mbox{\boldmath$\rm #1$}}

\usepackage{natbib}
\newcommand\eg{{e.g.}} 
\newcommand\etal{{et al.}} 

\newcommand\s{{\rm s}} 
\newcommand\yr{{\rm yr}} 
\newcommand\hr{{\rm hr}}
\newcommand\GHz{{\rm GHz}} 
\newcommand\m{{\rm m}} 
\newcommand\mm{{\rm m}\m} 
\newcommand\km{{\rm k}\m} 
\newcommand\mas{{\rm mas}} 
\newcommand\muas{\mu{\rm as}} 

\newcommand\e{{\rm e}}
\newcommand\sinc{{\rm sinc}}

\newcommand\SgrA{Sgr~A*}
\newcommand\EGS{J1745$-$2820}

\newcommand\kms{km~s$^{-1}$}

\newcommand\Msun{M$_\odot$}
\newcommand\Ro{R_0}
\newcommand\hho{H$_2$O}

\newbox\grsign \setbox\grsign=\hbox{$>$} \newdimen\grdimen \grdimen=\ht\grsign
\newbox\laxbox \newbox\gaxbox
\setbox\gaxbox=\hbox{\raise.5ex\hbox{$>$}\llap
     {\lower.5ex\hbox{$\sim$}}}\ht1=\grdimen\dp1=0pt
\setbox\laxbox=\hbox{\raise.5ex\hbox{$<$}\llap
     {\lower.5ex\hbox{$\sim$}}}\ht2=\grdimen\dp2=0pt

\begin{document}

\title{Limits on the Position Wander of Sgr A*}

\author{Mark J. Reid$^1$, Avery E. Broderick$^{1,2}$, Abraham Loeb$^1$,
        Mareki Honma$^3$, Andreas Brunthaler$^4$}
\affil{1. Harvard-Smithsonian Center for Astrophysics, 60 Garden Street, 
          Cambridge, MA 02138, USA\\
       2. Canadian Institute for Theoretical Astrophysics, 60 St. George Street, 
          Toronto, ON Canada\\
       3. Mizusawa VERA Observatory, NAOJ, Mitaka, Tokyo 181-8588, Japan\\
       4. Max-Planck-Institute f\"ur Radioastronomie, 69 Auf dem H\"ugel, 
          Bonn-53121, Germany}
\shorttitle{}
\shortauthors{Reid et al.}

\begin{abstract}
We present measurements with the VLBA of the variability in the
centroid position of \SgrA\ relative to a background quasar at $7\,\mm$
wavelength.  We find an average centroid wander of $71\pm 45\,\muas$ for
time scales between 50 and  $100\,\min$ and $113\pm50\,\muas$ for timescales
between 100 and $200\,\min$, with no secular trend.  These are
sufficient to begin constraining the viability of the accretion
hot-spot model for the radio variability of \SgrA.  It is possible to 
rule out hot spots with orbital radii above $15\,G M_{\rm Sgr A*}/c^2$ 
that contribute more than 30\% of the total $7\,\mm$ flux.  However,
closer or less luminous hot spots remain unconstrained.  Since
the fractional variability of \SgrA\ during our observations was 
$\sim20$\% on time scales of hours, the hot-spot model for \SgrA's 
radio variability remains consistent with these limits.  
Improved monitoring of \SgrA's centroid position has the 
potential to place significant constraints upon the existence and 
morphology of inhomogeneities in a supermassive black hole accretion flow.
\end{abstract}

\keywords{accretion disks -- astrometry -- black hole physics -- 
Galaxy: center -- gravitational lensing -- 
instrumentation: high angular resolution}

\maketitle

\section{Introduction}
There is now overwhelming evidence that \SgrA\ is a supermassive
black hole at the center of the Milky Way.  Many stars are
observed to orbit about a common focal position, requiring 
an unseen mass of $\approx4\times10^6$~\Msun\ contained within a 
radius of less than 100~AU \citep{Schoedel:2002,Ghez:2003}, 
for a distance to the center of 8.0~kpc \citep{Reid:1993}.  
Accurate registration of the infrared and radio reference frames 
\citep{Menten:1997,Reid:2003} reveal that the common
orbital focal position is coincident with \SgrA\ to within
measurement uncertainty of $\approx10$~mas.  Finally, the
absence of intrinsic motion of \SgrA\ at levels near that 
expected for a $4\times10^6$~\Msun\ object 
\citep{Reid-Brun:2004}, coupled with a size less than
~1~AU \citep{Bower:2004}, provide a lower limit on mass
density of $\sim10^{22}$~\Msun~pc$^{-3}$, which is only two orders of 
magnitude less than the density of a $4\times10^6$~\Msun\ non-rotating  
black hole within its innermost stable orbit.  There can now be little
doubt that \SgrA\ is a supermassive black hole.

\citet{Reid-Brun:2004} present measurements of
the position of \SgrA\ relative to a compact extragalactic radio source
(\EGS, also refered to as J1745-283 in earlier publications).  
These measurements were conducted with the NRAO
\footnote{The National Radio Astronomy Observatory is operated by Associated 
Universities, Inc., under a cooperative agreement with the National Science 
Foundation.}
Very Long Baseline Array (VLBA) over a period of $8\,\yr$ at a wavelength of 
$7\,\mm$ ($43\,\GHz$) and have been used to determine the
apparent proper motion of \SgrA.  Over time scales of months or longer, 
\SgrA's {\it apparent} motion is dominated by the 
effects of the orbit of the Sun about the center of the Galaxy. 
The component of the Sun's orbit in the Galactic plane is uncertain
at roughly the 10\% level, and this limits estimation of any intrinsic
motion of \SgrA\ at about the $\pm20$~\kms\ level.  However, the component of the
motion of the Sun out of the Galactic plane is known to high accuracy
[$7.16\pm0.38$~\kms\ toward the North Galactic Pole; \citet{Dehnen-Binney:1998}].
After removing the effects of the Sun's motion, the residual motion of 
\SgrA\ perpendicular to the Galactic plane is very small,
$\lesssim 1\,\km\,\s^{-1}$,
as expected for a supermassive black hole (SMBH) at the dynamical center of
the Galaxy.
While previous work concentrated on the long-term motion of 
\SgrA, here we analyze its short-term position ``wander'' on time scales 
of hours to weeks.  

Short-timescale motion of the centroid position of \SgrA\
would be expected if a portion of the emission comes from material
orbiting about the SMBH.  The degree of centroid variability would
necessarily depend upon the brightness of the orbiting material, the
degree to which its emission is nonuniform, and the orbital radius
dominating the total flux.  We use a simple hot-spot model to relate the
constraint from the observed short-term position wander of
\SgrA\ (\S \ref{OoCM}) to a constraint upon the presence of strong 
inhomogeneities in the accretion flow onto the SMBH as a function of 
hot-spot luminosity and orbital period (\S \ref{Constraints}).  
Finally, concluding remarks are contained in \S \ref{C}.

\section{Observations of Centroid Motion} \label{OoCM}
\citet{Reid-Brun:2004} describe the observations and data calibration 
methods in detail.  Briefly, we obtained position data as follows:
the VLBA antennas switched between \SgrA\ and a compact 
extragalactic source (\EGS) every 15 seconds in order provide 
interferometer phase differences rapidly enough to cancel the 
effects of short term atmospheric fluctuations.  
The stronger source, \SgrA, was used as
the phase reference to calibrate data from the weaker source, \EGS.

Astrometric imaging of \SgrA\ at 7-mm wavelength is best
accomplished with only the five inner-VLBA antennas (FD, KP, LA, OV 
and PT).  These antennas produce interferometer baselines with 
lengths of up to $1500$~km, resulting in synthesized beams
typically about $2.5 \times 0.7$~mas (FWHM) elongated north-south.
Longer baselines (\eg\ involving the Washington (BR) 
and Iowa (NL) state antennas) 
are not generally useful for precise astrometry, as it is difficult to 
detect \SgrA\ with the 8-sec on-source integrations afforded by
rapid switching, coupled with low fringe visibilities on long baselines
owing to the large, scatter-broadened, image of \SgrA.  Also, the sources 
are mutually visible with the inner five antennas for only a short time 
period for antennas far from the inner ones.

Our most accurate astrometry was obtained with atmospheric path-delay 
calibration using ``geodetic'' blocks \citep{Reid-Brun:2004}.
This involves short periods of observations of quasars with
a wide spanned-bandwidth and scheduled to deliver
a wide range of source elevations.  These geodetic blocks were placed
before the start, at the middle, and after the end of the \SgrA\ 
observations.  Analysis of these data yield estimates of the zenith 
atmospheric path-delay at each antenna accurate to $\sim0.5$ to 1~cm 
(or about 1 wavelength).

\subsection{Position wander: hours}\label{section:hours}

For analysis of short-term wander, we selected only our highest 
quality data, requiring both high accuracy atmospheric path-delay 
calibration using geodetic blocks (which was started in 2003) and
data from all five inner-VLBA antennas.  
Data from VLBA programs BR84 on 2003 April 5 and 25 and 
BR124 on 2007 April 5 and 11 satisfied these requirements. 

\begin{figure*}
\begin{center}
\begin{tabular}{cc}
\includegraphics[width=0.75\textwidth]{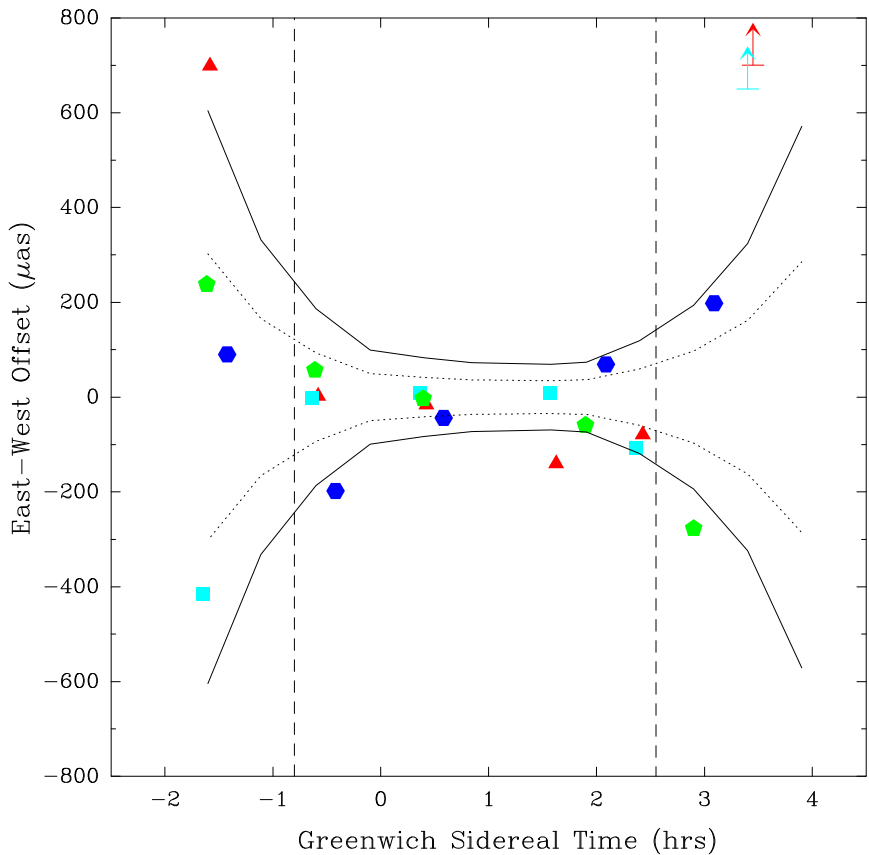}
\\
\end{tabular}
\end{center}
\caption{Hourly measurements of the centroid position of Sgr~A*
relative to an extragalactic source.  Data from 2003 April 05
({\it red triangles}) and 25 ({\it cyan squares}) and 2007 April 05
({\it green pentagons}) and 11 ({\it blue hexagons}) are plotted.
All positions are relative to their daily average position.
{\it Dotted} and {\it solid lines} indicate expected $\approx1\sigma$ 
and $\approx2\sigma$
uncertainties, based on Monte Carlo simulations of the effects
of uncompensated atmospheric path-delays (0.5~cm rms) at each antenna.
Data outside the vertical {\it dashed lines} were excluded from analysis
owing to their large expected position errors.
         \label{hourly_positions}
        }
\end{figure*}

The position wander of \SgrA\ over time scales of hours was determined 
by dividing the calibrated interferometer data into hourly bins.
The data were Fourier transformed to make images and intensity
centroid positions were determined.  
In practice, we measured the background source, \EGS, which had
been phase-referenced to \SgrA, but we interpret any position changes
as owing to changes in \SgrA.
In Fig.~\ref{hourly_positions} we show the East-West (EW) position 
offsets as a function of Greenwich sidereal time, 
after removing an average position for each day's data.  
The North-South (NS) positions are intrinsically less accurate by a 
factor  of about 3, as the NS projections of the interferometer baselines
are correspondingly shorter than the EW projections.

At the start and end of the observing tracks, the sources are
at low elevations and susceptible to large interferometer phase
shifts caused by uncompensated atmospheric path-delays.  We 
estimated the effects of zenith path-delay errors with 1000 
independent simulations. 
Each simulation started by randomly selecting zenith path-delay
errors (0.5~cm rms) for each antenna.  We then calculated the expected 
interferometer phase shifts, dependent on time-varying source
zenith angles, at half-hour time steps.  Position shifts
were estimated as the product of the phase shifts (in turns)
multiplied by an approximate projected interferometer fringe 
spacing.  Finally, we calculated a weighted position shift for all
baselines and all times when the source was above $15^\circ$ elevation.
The resulting expected $\pm1\sigma$ and $\pm2\sigma$ position 
error envelopes are plotted in Fig.~\ref{hourly_positions}.
Based on the observed and simulated increase in position scatter for
Greenwich sidereal times (GST) before $-0.8$~hr and after $+2.6$~hr,
we only use data within this time range for our study of hourly
position wander. 

\begin{figure*}
\begin{center}
\begin{tabular}{cc}
\includegraphics[width=0.65\textwidth]{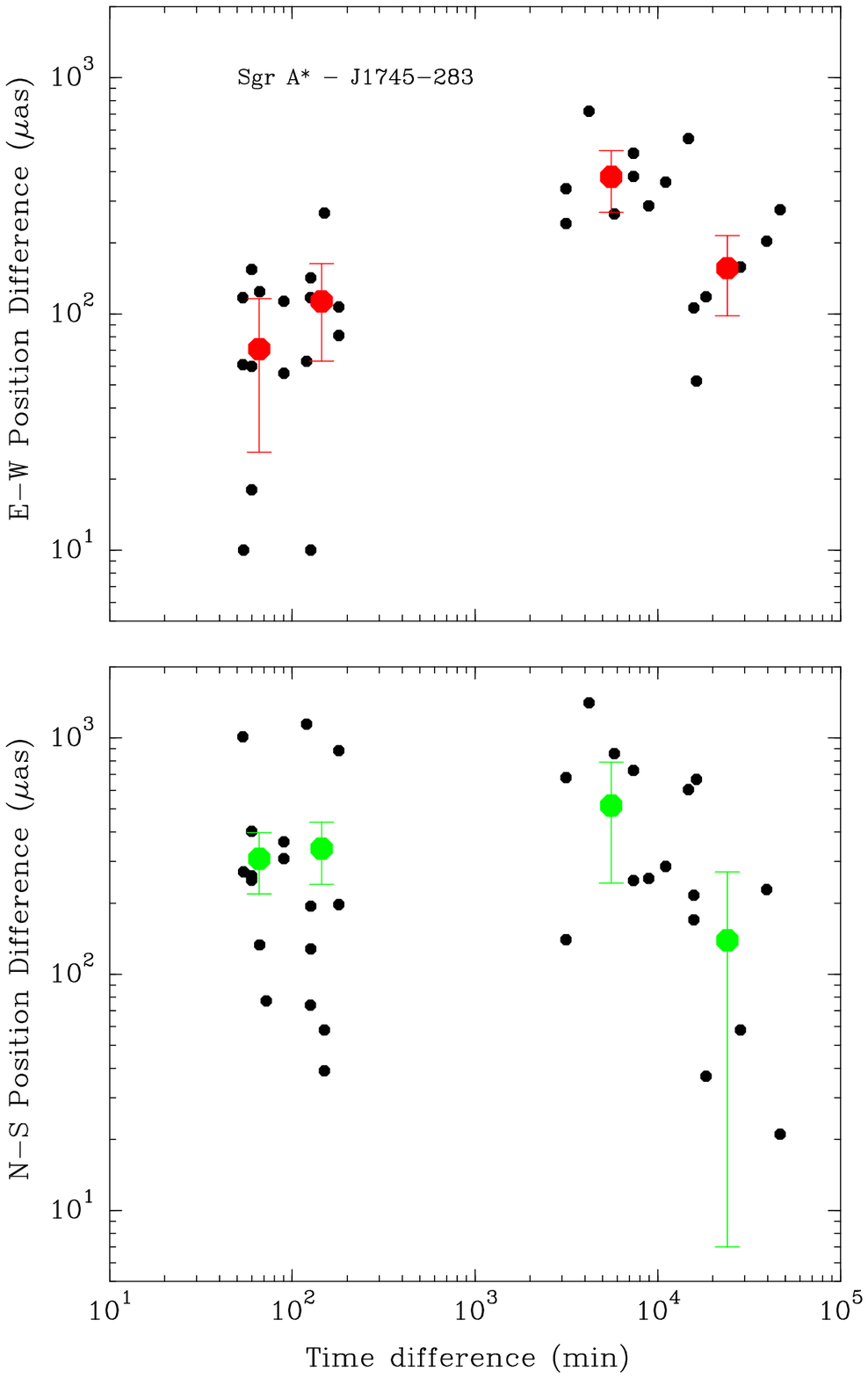}
\\
\end{tabular}
\end{center}
\caption{Position wander of \SgrA.  Plotted are East-West ({\it top panel}) 
and North-South ({\it bottom panel}) position differences as a function of
time separation.  Also shown are binned average position differences
as large points with error bars.  These position differences are likely
caused by uncompensated atmospheric path-delay errors and should be 
considered as upper limits to the position wander of \SgrA.
         \label{wander}
        }
\end{figure*}

We differenced the EW and NS positions 
for all pairs of measurements separated by less than $10^3$~min; 
the magnitudes of these position differences are shown on the left-hand 
side of Fig.~\ref{wander}, along with 2 binned weighted averages.  
The average difference in position for measurements separated by 50 to
$100$~min is $71\pm45~\muas$ and for measurements separated
by 100 to 200~min is $113\pm50~\muas$.
Neither average position difference deviates significantly from
zero, and the position differences are consistent with systematic errors
owing predominantly to mis-modeling zenith propagation path-delays through 
the Earth's atmosphere at the $\approx1$~cm level.
Thus, we use these measurements as upper limits for the 
position wander of the centroid of \SgrA's emission.

\subsection{Position wander: days to weeks}
The position wander of \SgrA\ on time scales of days to weeks was
determined from ``daily'' position measurements, obtained from
$\approx5$ hours of data when source elevations were above 
$\approx20^\circ$ at most of the five inner-VLBA antennas.  These positions
were plotted in Fig.~3 of \citet{Reid-Brun:2004}, after removing
the long-term proper motion of \SgrA\ of $6.379$~\mas~yr$^{-1}$
along a position angle of $209.60^\circ$ East of North.
We differenced the EW and NS positions 
for all pairs of measurements separated by less than $10^5$~min; 
the magnitudes of these position differences are shown in Fig.~\ref{wander},
along with weighted averages for two time-bins.

The average difference in EW position for measurements separated by
$\approx6000$~min ($\approx4$~days) is about $400~\muas$.
This difference is likely not a property of the emission of \SgrA;
instead it is probably caused by mis-modeling large-scale
propagation delays through the Earth's atmosphere.  
Most of the data used in this analysis were collected before 
we started to measure atmospheric path-lengths in 2003.
Without this calibration, residual zenith path-length errors of $\sim5$~cm 
are typical for the VLBA correlator model at 7-mm wavelength.
Since propagation delays can be correlated with large-scale weather patterns,
which have characteristic time scales of several days, one 
expects to see such a residual signature in our position differences. 

\section{Constraints upon hot-spot models of Sgr A* variability} \label{Constraints}
Limits upon the variability in the centroid position of high resolution
images of Sgr A* imply a constraint upon orbiting hot-spot models
for Sgr A*'s radio variability.  This will necessarily be a function
of the hot-spot orbit and the hot-spot/disk flux ratio.  In this
section we derive the expected variability for two simple hot-spot
models.  The first is an optically thin Newtonian hot-spot, while the
second fully incorporates general relativity and the opacity of a disk
constructed such that it reproduces the observed spectrum of Sgr A*.

\subsection{Idealized Newtonian hot-spot centroid variability}
For the idealized Newtonian hot-spot we assume that there is no
lensing by the black hole, the disk is completely optically thin, and
the hot-spot is a point (the first two are physically appropriate for
spots at large orbital radii).  In this case the hot spot's orbit
(which we assume to be circular with radius $r$) is simply given by 
Kepler's law:
\begin{equation} 
x(t) = r \cos(\Omega t + \phi)\,\quad
y(t) = r \cos i \sin(\Omega t + \phi)\,
\end{equation}
where $i$ is the orbital inclination, 
\begin{equation}
\Omega = \sqrt{\frac{GM_{\rm Sgr A*}}{r^3}}\,,
\end{equation}
$M_{\rm Sgr A*}$ is the mass of Sgr A*, and $\phi$ is an arbitrary phase.

The image centroid, $\bmath{X}_C$ is constructed by integrating the
source emission over a time $T$, which need not be small in comparison
to the orbital period, and thus the motion of the hot-spot will
generally be important.  Explicitly, if we set the centroid of the
disk emission to be at the origin,
\begin{equation}
\bmath{X}_C = 
\frac{\int_0^T F_{\rm spot}(t) \bmath{x}(t) dt}
{\int_0^T \left[ F_{\rm disk} + F_{\rm spot}(t) \right] dt}\,.
\label{eq:avg_cent}
\end{equation}
This will generally be a function of the initial phase of the orbit,
$\phi$, and the integration time $T$.  If we make the simplifying
assumption that the hot-spot flux may be treated as roughly constant,
then this reduces to
\begin{equation}
\bmath{X}_C = 
\eta
\overline{\bmath{x}}(\phi;T)
\quad\text{where}\quad
\eta \equiv \frac{F_{\rm spot}}{F_{\rm disk} + F_{\rm spot}}
\,,
\end{equation}
and $\overline{x}(\phi;T)$ is the average spot position over some
time $T$ with initial orbital phase\footnote{While this
  assumption may appear to be manifestly unjustified, for an optically
thick hot-spot in the Rayleigh-Jeans limit, the final expression is
formally accurate!  The reason is that the relativistic beaming of the
hot-spot emission is precisely countered in this instance by the
combination of the Doppler-boosting and time-of-flight delays
\citep{Brod-Loeb:2005}.} $\phi$.  It is straightforward to show that
\begin{eqnarray}
\displaystyle \overline{x}(\phi;T)
&=&
\displaystyle r~\sinc\left(\frac{\Omega T}{2}\right) \cos\phi \nonumber \\
\displaystyle \overline{y}(\phi;T)
&=&
\displaystyle r~\sinc\left(\frac{\Omega T}{2}\right) \cos i \sin\phi\,,
\end{eqnarray}
where a factor of $\Omega T/2$ has been subsumed into the arbitrary
phase $\phi$.  This corresponds to a single observation of the
position of \SgrA.

To study the position wander, we must compare two such measurements
separated by some time $\Delta T$.  This results in a change in the
centroid position of
\begin{eqnarray}
&&\displaystyle \left|\Delta \bmath{X}_C \right|(\phi;T,\Delta T)
=
\eta
\left|\overline{\bmath{x}}(\phi+\Omega\Delta T;T)-\overline{\bmath{x}}(\phi;T)\right| \nonumber\\
&&\qquad=
\displaystyle \eta~
r~\sinc\left(\frac{\Omega T}{2}\right)
\sin\left(\frac{\Omega\Delta T}{2}\right) \nonumber \\
&&\qquad\qquad\qquad\displaystyle\times\sqrt{\cos^2\phi + \cos^2 i \sin^2 \phi}\,.
\end{eqnarray}
Maximizing the position wander over the arbitrary initial phase simply
removes the terms in the radical:
\begin{equation}
\left| \Delta \bmath{X}_{C,\rm max} \right| (T,\Delta T)
=
\eta~r~\sinc\left(\frac{\Omega T}{2}\right) \sin\left(\frac{\Omega\Delta T}{2}\right)\,.
\label{eq:dr1D}
\end{equation}
While it may be unlikely that any given observation of the position
wander will allow detection of the maximum centroid displacement, 
this quantity places a strict constraint upon hot-spots.  
Alternatively, one may wish
to compare the observations to the predicted RMS centroid variability,
which in this case is roughly $70\%$ of
$\left| \Delta \bmath{X}_{C,\rm max} \right| (T,\Delta T)$.

For fixed $T$ and $\Delta T$,
$\left|\Delta\bmath{X}_{C,\rm max}\right|$ vanishes when an integral
number of periods is commensurate with either the integration time
(and thus $\overline{\bmath{x}}(\phi;T)$ vanishes identically) or the
delay between measurements.  These are evident in Figure
\ref{fig:drs}, in which the thin blue line shows 
$\left|\Delta\bmath{X}_{C,\rm max}\right|$ as a function of orbital
period for a number of values of $\eta$ with $T=1\,\hr$ and $\Delta
T=4\,\hr$.  However, by choosing many different $T$ and $\Delta T$, it
is possible fill in the nulls associated with each individual
choice.  There will be constraints upon what values of $T$ and $\Delta
T$ are possible, e.g., due to sensitivity and imaging requirements
(see \S \ref{OoCM}).  The upper-envelope of the observed centroid
displacement for $1\,\hr \le T \le 4\,\hr$ and
$1\,\hr \le \Delta T \le 4\,\hr$ is shown in Fig. \ref{fig:drs} by the
thick, solid blue line.   For reference we also show, with dotted 
blue lines, the maximum deviation possible, corresponding to the 
case when all of the flux is due to the hot spot.

Generally, the observed displacement has a maximum near orbital
periods of roughly $\pi\sqrt{T\Delta T}$.  That such a maximum must
exist can be inferred from the facts that (i) at large orbital radii
the Keplerian velocity decreases and thus beyond some distance the hot
spot will move sufficiently slow as to produce no detectable change in
it's position over the time $\Delta T$, and (ii) at small orbital
radii the intrinsic variability in the spot position is smaller and
the hot-spot makes many complete orbits in the integration time
$T$, and thus the variable portion of the centroid position is
dominated by a small fraction of the integrated flux, 
yielding again a small observable change in the centroid position.  
However, we may also show this explicitly by considering the
asymptotic expansions of eq. (\ref{eq:dr1D}).  For small $\Omega$
(large period $P$), the $\sinc$-term is roughly unity and
\begin{equation}
\left| \Delta \bmath{X}_{C,\rm max} \right| (T,\Delta T)
=
\eta r \frac{\Omega \Delta T}{2}
\propto
r^{-1/2}
\,.
\end{equation}
In contrast, for large $\Omega$, 
$\left| \Delta \bmath{X}_{C,\rm max} \right| (T,\Delta T)$ is
strongly oscillatory.  Nevertheless, it is bounded from above by the
$\sinc$ term, and thus
\begin{equation}
\left| \Delta \bmath{X}_{C,\rm max} \right| (T,\Delta T)
\le
\frac{2 \eta r}{\Omega T}
\propto
r^{5/2}
\,.
\end{equation}
Setting these limiting expressions equal to each other gives the
desired condition that a maximum observed displacement occurs near
$\Omega \simeq 2/\sqrt{T\Delta T}$.  Note that this is true for the
envelope obtained by varying $T$ and $\Delta T$ in the prescribed
ranges as well, and serves as a simple estimate of the sensitivity of
these types of measurements.

\subsection{Fully relativistic, hot-spot in an optically thick disk}
We now consider a more realistic model in which a hot spot is
embedded in an accretion disk, including the relativistic beaming,
Doppler boosting, strong gravitational lensing and the opacity of the
disk and hot spot.  This is necessarily a more complicated model, and
thus we address it numerically via the ray-tracing, radiative-transfer
code described in \citet{Brod-Loeb:2006} (to which we direct the
reader for more information, the model only being summarized below).

Due to its ability to shield the hot spot from view, the structure of
the background disk is of particular importance.  In the absence of an
unambiguous prediction from existing accretion flow theory, we have
modeled it as a self-similar Radiatively Inefficient Accretion Flow
(RIAF) following \citet{Yuan-Quat-Nara:2003}.  Specifically, the
accretion flow is characterized by a Keplerian velocity distribution,
a population of thermal electrons with density and temperature
\begin{equation}
n_{e,\rm th} = n^0_{e,\rm th} r^{-1.1} \e^{-z^2/2\rho^2}
\quad\text{and}\quad
T_{e} = T^0_{e} r^{-0.84}\,,
\end{equation}
respectively, a population of non-thermal electrons
\begin{equation}
n_{e,\rm nth} = n^0_{e,\rm nth} r^{-2.9} \e^{-z^2/2\rho^2}\,,
\end{equation}
and spectral index $\alpha_{\rm disk} = 1.25$ 
(defined as $S\propto\nu^{-\alpha}$),
and a toroidal magnetic field in approximate ($\beta=10$) equipartition
with the ions (which produce the majority of the pressure), i.e.,
\begin{equation}
\frac{B^2}{8\pi}
=
\beta^{-1} n_{e,\rm th} \frac{m_p c^2}{6 r}\,.
\end{equation}
In all expressions $r$, $\rho$ and $z$ are measured in units of
$G M_{\rm Sgr A*}/c^2$.  The power-laws are taken from
\citet{Yuan-Quat-Nara:2003},
and the three coefficients ($n^0_{e,\rm th}$, $T^0_{e}$ and
$n^0_{e,\rm nth}$) are set by fitting the radio, sub-mm and
near-infrared spectrum of Sgr A*.  While this is not a unique model
for the accretion flow around Sgr A*, it is representative of the
general class of RIAFs, and we expect that our results will be quite
generic.

\begin{figure*}
\begin{center}
\begin{tabular}{cc}
\includegraphics[width=0.4\textwidth]{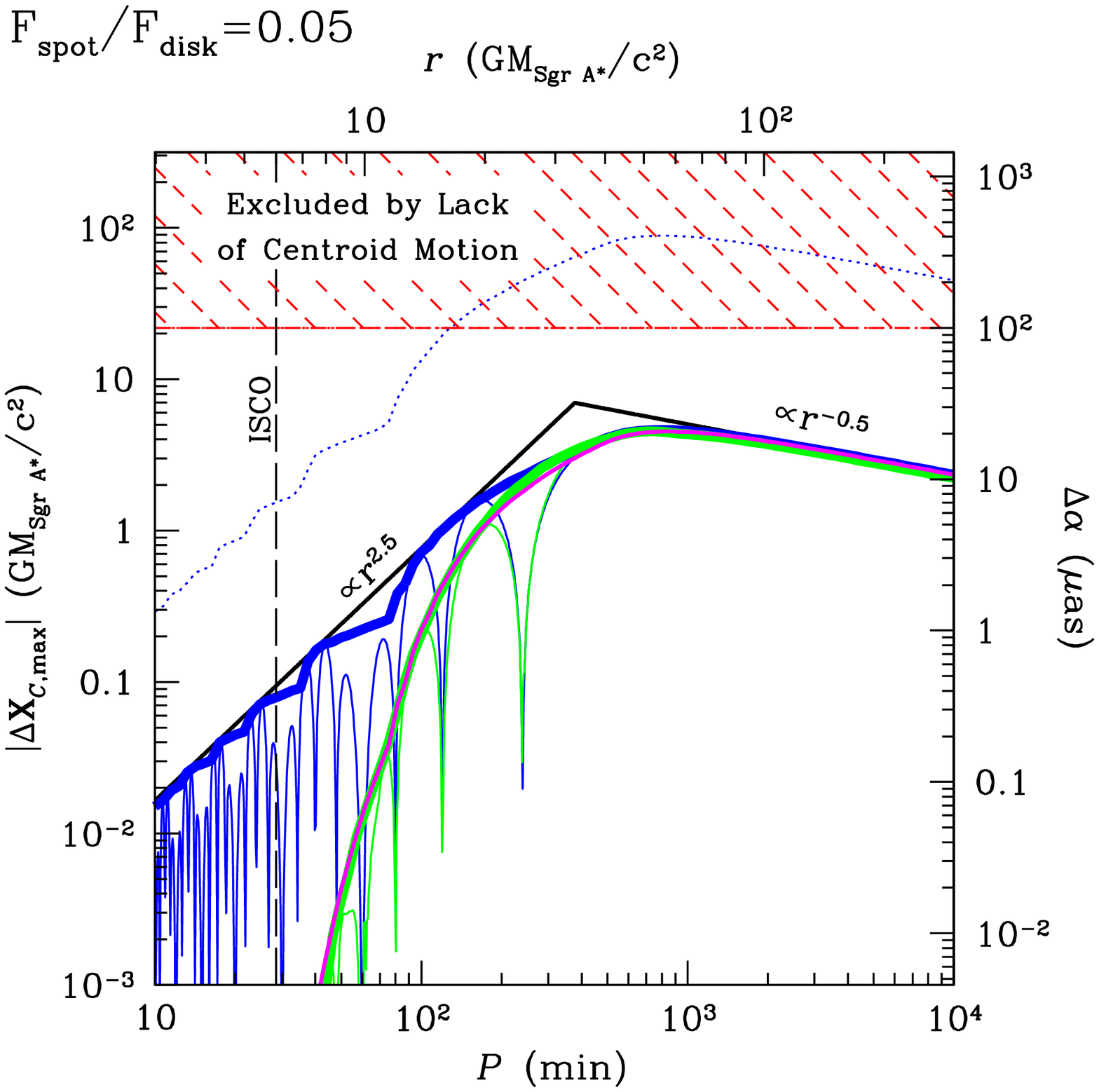}
&
\includegraphics[width=0.4\textwidth]{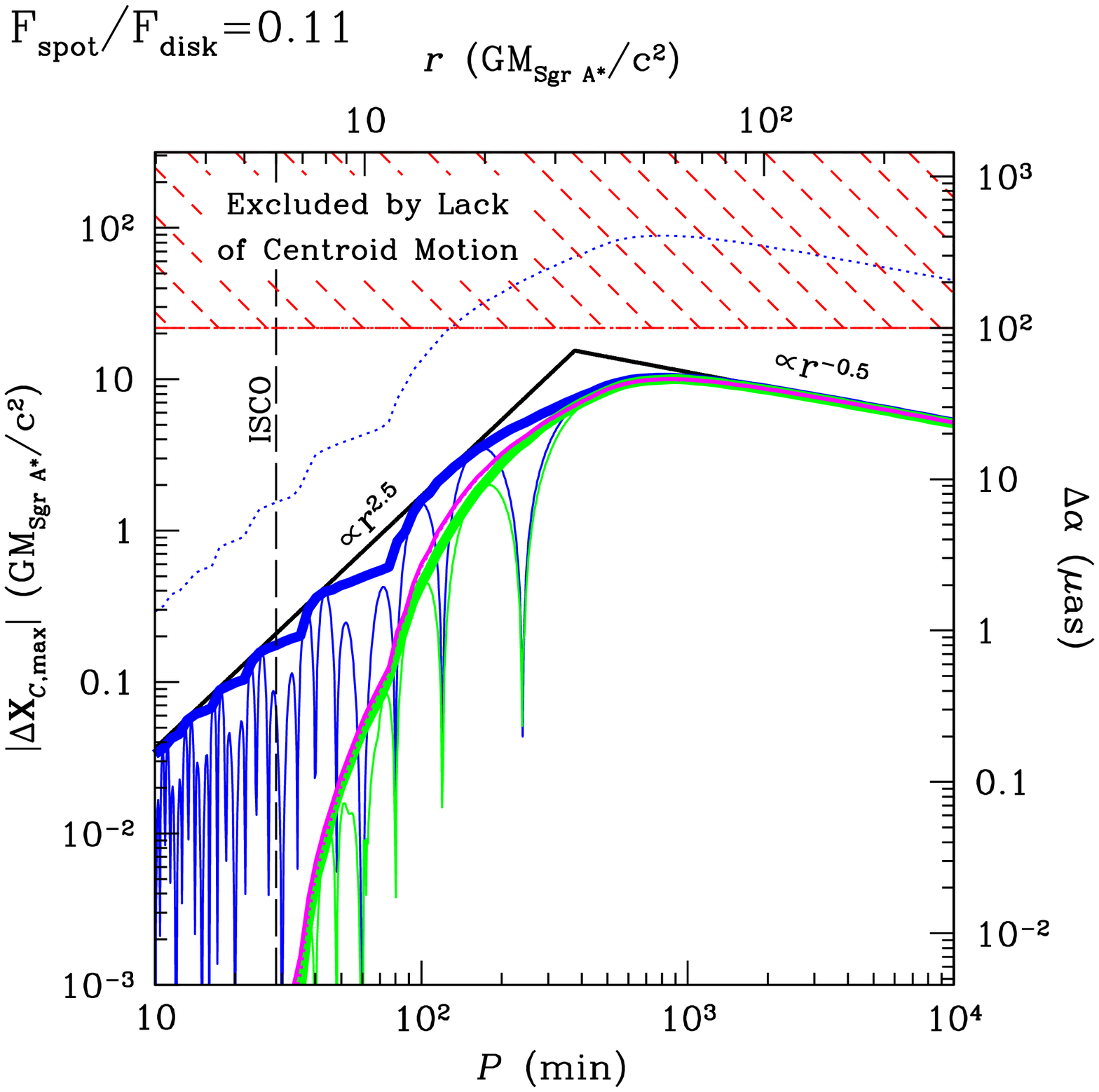}\\
\includegraphics[width=0.4\textwidth]{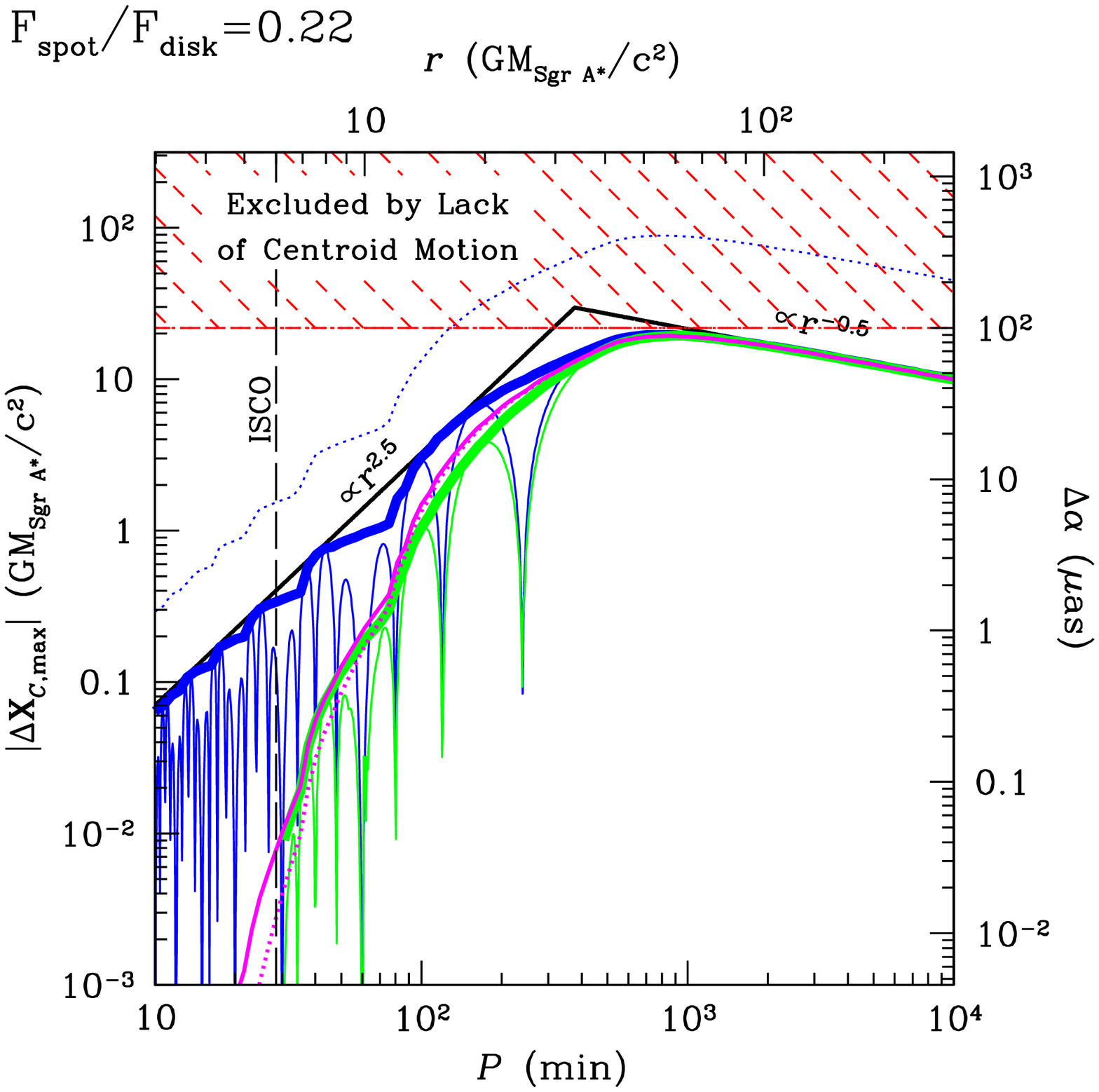}
&
\includegraphics[width=0.4\textwidth]{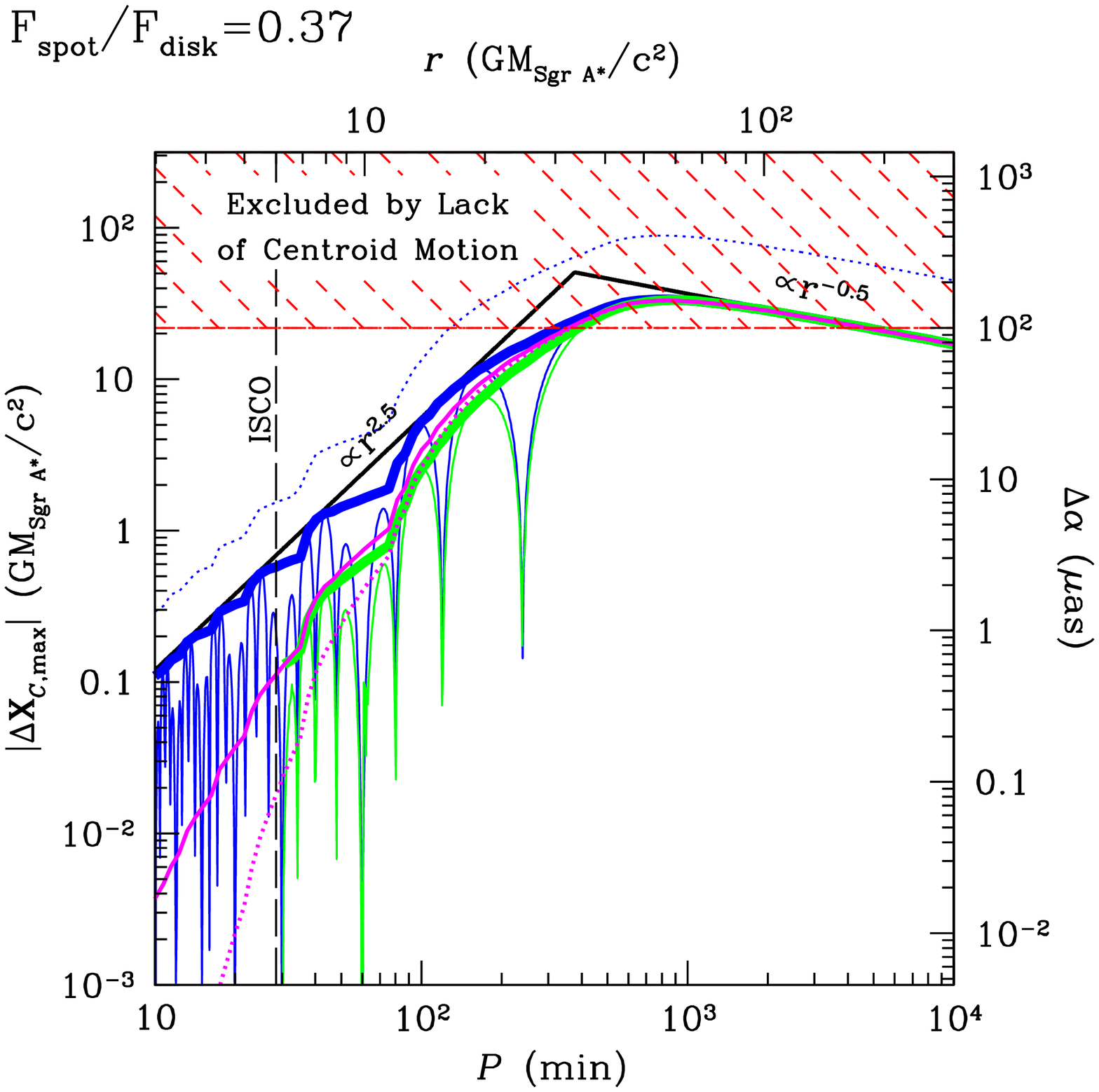}\\
\end{tabular}
\end{center}
\caption{Maximum observed physical displacement of the centroid as a
  function of the orbital period for the idealized Newtonian case
  (blue) and the fully general relativistic case appropriate for Sgr
  A* (green) for a number of spot-disk flux ratios.
  The period and angular offset scales assume a distance of 8.0 kpc and 
  M$_{Sgr A*} = 4 \times 10^6$~\Msun.
  The thin blue and green lines show the variation associated with an
  observations integrated over $T=1\,\hr$ and separated by $\Delta
  T=4\,\hr$.  The thick blue and green lines show the envelope for all
  observation strategies with $T\ge 1\,\hr$ and $\Delta T\le 4$.  In
  all cases the thin dotted blue line shows the maximum deviation
  possible, corresponding to when all of the flux is due to the hot
  spot.  The
  solid black lines show the asymptotic power-law behavior at both
  short and long periods.  The dotted and solid magenta lines shows
  the opacity modified idealized Newtonian case without and with
  lensing, respectively.  Finally, the regions excluded by $7\,\mm$ VLBI
  are shown by the red hatched regions.  For convenience the orbital
  radii and angular scale of the centroid displacements are shown on
  the top and right axes, respectively. \label{fig:drs}}
\end{figure*}

The hot-spot is modeled by a spherical (in it's frame) Gaussian
over-density of non-thermal electrons, with 
a radial scale of $1.5 G M_{\rm Sgr A*}/c^2$.  The
primary distinction between the systems shown in the panels of
Fig. \ref{fig:drs} is the size of the photosphere of the hot-spot
($\delta r$), larger hot-spots necessarily being more luminous.

The radiative transfer is assumed to be dominated by synchrotron
emission and absorption.  We follow \citet{Jone-ODell:1977} for the power-law
electron distributions in the accretion flow and the hot spot, cutting
the electron distributions off below Lorentz factors of $10^2$,
roughly in agreement with the assumptions in
\citet{Yuan-Quat-Nara:2003}.  We treat the radiative transfer of the
thermal component of the accretion flow in a similar fashion as that
discussed in \citet{Yuan-Quat-Nara:2003}, appropriately altered to
account for the relativistic nature of the bulk motion (see, e.g.,
\citealt{Brod-Blan:2004} for more detail on how this may be done).

The centroid displacements were computed by (i) generating a
sequence of images associated with an entire orbit, (ii) determining
the instantaneous image centroid for each, (iii) inserting these into
eq.~(\ref{eq:avg_cent}) to determine the time-integrated centroid
position for a given $T$, and finally (iv) given a $\Delta T$ varying
the orbital phase until the maximum
$\left| \Delta \bmath{X}_{C,\rm max} \right| (T,\Delta T)$
was obtained.  The result for $T=1\,\hr$ and $\Delta T=4\,\hr$ is
shown by the thin green line in Fig. \ref{fig:drs}, and exhibits
many of the features found in the idealized Newtonian case.  This
procedure was repeated a number of times for randomly chosen
$T \in [1\,\hr,4\,\hr]$ and $\Delta T \in [1\,\hr,4\,\hr]$  to produce
the upper-envelope of the centroid displacements, and is shown by the
thick green line in Fig. \ref{fig:drs}.

The primary effect is the suppression of the centroid variability due
to the opacity of the background accretion flow.  This suppression
becomes substantial at radii near the location of the photosphere,
which is approximately $12.5 G M_{\rm Sgr A*}/c^2$ at $7\,\mm$ wavelength.
In contrast, at large orbital radii the idealized Newtonian model fits
quite well.  This interpretation is further supported by the fact that the
idealized Newtonian model fits quite well once it has been modified to
include opacity.  Specifically, as a rough approximation, we reduce the
hot-spot flux by a factor of $\e^{-\tau}$, where
\begin{equation}
\tau = \left(\frac{r+\delta r}{12.5}\right)^{-2.99}
\propto r n_{e,\rm nth} B^{\alpha_{\rm disk} + 3/2}
\end{equation}
is the optical depth associated with the non-thermal electrons in the
accretion flow.  This is shown by the magenta lines in
Fig. \ref{fig:drs}.  While we should also include the thermal
electron component, at $7\,\mm$~wavelength the non-thermal electrons appear to
dominate the opacity.

The free parameter $\delta r$ is associated with the finite extent of
the hot spot.  That is, it is possible for the spot to be visible even
if the hot-spot center is inside of the accretion-flow photosphere.
It is for this reason that the centroid variability is more similar to
the idealized Newtonian value for high-luminosity spots than for
low-luminosity spots.  Setting $\delta r$ using the hot-spot
photosphere radius alone gives the dotted magenta line, which
underestimates the centroid variability considerably at small radii
for large (bright) hot spots.  This is due to the failure of the
idealized Newtonian calculation to account for the strong lensing of
large spots in small orbits (with $r_{\rm spot}\simeq r$).  However,
simply employing a larger $\delta r$ (64\% larger at the largest
disk/hot-spot flux ratio we consider) results in a substantially
improved fit.

\section{Discussion} \label{C}
Possible reasons for position wander include intrinsic
variations in the position of the emitting plasma
(\eg, variations in the accretion flow or perhaps in a jet)
or extrinsic processes such as refractive interstellar scattering. 
\SgrA\ is observed to be diffractively scattered to a
size of $\theta_{sc} \sim 0.5 (\lambda/0.7~{\rm cm})^2$~mas,
where $\lambda$ is the observing wavelength.
Flux density fluctuations are modest and decrease
in strength with increasing wavelength; thus strong refractive 
scintillations are not indicated \citep{Gwinn:1991}.  
Any refractive position wander should be 
$\ll \theta_{sc}$ and should occur on time scales
$>\theta_{sc}D/v$, where $D$ is the distance and $v$ 
is the transverse velocity of the scattering ``screen'' 
relative to the observer \citep{Romani:1986}.  
For $D\approx\Ro\approx8$~kpc \citep{Reid:1993} and 
$v\sim100$~\kms, characteristic of material in the inner 
$\sim100$~pc of the Galaxy where large scattering sizes 
are observed, the refractive time scale is $>10^3$~hours.  
Thus, we would not expect a significant contribution to the 
short-term wander of \SgrA\ from refractive scattering.  
For comparison, \citet{Gwinn:1988}, using VLBI observations 
of the Sgr~B2(N) \hho\ masers near the Galactic center, 
find a wander limit of $<18~\muas$ over timescales of months 
for maser spots, which are diffractively scattered to a 
comparable size (at 22 GHz) as \SgrA\ (at 43 GHz). 
Of course, our results provide an observation limit to any 
refractive position wander.

Since extrinsic sources of position wander (scattering) are 
unlikely to be dominant, we now discuss the implications for
intrinsic wander from variations in brightness within an accretion 
disk given in \S\ref{Constraints}.
Our observations of the lack of short-term wander of the centroid 
position of \SgrA\ presented in \S\ref{section:hours}
give an upper limit of $\approx100~\muas$ for time scales of
$\approx1~{\rm to}~4$ hours.  
This translates to an upper limit
on the wander versus orbital period plots in Fig.~\ref{fig:drs}
as indicated by the horizontal red line and  hatched region.  
(In the very unlikely event that the accretion disk inclination is both
near $90^\circ$ and oriented nearly North-South on the sky, we would need
to use our NS limits, which are a factor of three weaker.)
Our EW limit is below the dotted blue line in Fig.~\ref{fig:drs}, which is
for hot spot flux density dominating over disk (or possible
jet) emission, for orbital periods exceeding 120~min (corresponding
to orbital radii larger than $15\,G M_{\rm Sgr A*}/c^2$ for
$M_{\rm Sgr A*}= 4\times10^6$ \Msun).   For cases 
in which the hot spot flux density is weaker than that of the disk,
somewhat longer periods are allowed.  For example, for 
$F_{\rm spot}/F_{\rm disk} = 0.37$, orbital periods longer than 5~hr are
excluded.  

In practice, the limits placed by current $7\,\mm$ VLBI are 
significantly weaker.  The limited sensitivity to hot spots on 
compact orbits is primarily due to two reasons: 
(1) ``long'' integration times  ($T\gtrsim 1\,\hr$) average much 
of the short time variability out, and
(2) the opacity of the accretion flow itself makes it difficult 
to view hot-spots on compact orbits at $7\,\mm$.
The integration time is limited by the sensitivity issues and
the small number of antennas yielding interferometer baselines 
$<1500$~km afforded by the current VLBA; higher bandwidth recording 
in the future should help alleviate this problem. 
The optical depth is a property of Sgr A* itself, and can only be 
addressed by observations at shorter wavelengths.   
However, even in the absence of an optically thick accretion flow, 
it is not possible to increase the centroid variability by more than 
an order of magnitude due to the intrinsically small orbital radii, 
as seen by comparing the blue and green limits in the lower-left 
panel of Fig. \ref{fig:drs}.

Nevertheless, high-resolution astrometry is reaching sensitivities and
resolutions sufficient to begin to test the hot-spot model for bright
Sgr A* flares.  Unfortunately, the typical fractional variability at
$7\,\mm$ during our observations was roughly $\pm20\%$, implying
that significant improvement in positional accuracy will be required
to constrain such events.  Since the observed centroid
wander is consistent with systematic errors, owing predominantly to
centimeter-scale errors in the modeling of the atmospheric
path-delays, substantially increasing accuracy
will require better calibration techniques.   
However, for the somewhat rare instances in which
the spot is substantially brighter \citep{Zhao:2001}, the VLBA at
7, or possibly 3,~mm wavelength appears poised to provide significant 
limits upon the existence and morphology of inhomogeneities in the 
accretion flow surrounding \SgrA. 

Ultimately, observations at $\sim1$~mm wavelength with VLBI techniques 
or at infrared wavelengths with an instrument like GRAVITY 
\citep{GRAVITY} may be necessary to image the region within 
$\sim3$ Schwarzschild radii on the short time scales needed to test 
the hot-spot model.

\vskip 1truecm
A.B. is supported by the Priority Programme 1177 of
the Deutsche Forschungsgemeinschaft.
\vskip 0.5truecm
{\it Facilities:} \facility{VLBA}

\end{document}